\documentclass[aps,prb,amssymb,twocolumn]{revtex4}

\newcommand{\bfq}{{\bf q}}
\newcommand{\bfp}{{\bf p}}
\newcommand{\bfk}{{\bf k}}
\newcommand{\bfA}{{\bf A}}
\newcommand{\bfr}{{\bf r}}
\newcommand{\bfx}{{\bf x}}

\begin{document}
\draft

\title{Collective Modes and Raman Scattering in One Dimensional Electron
Systems}

\author{D.-W. Wang$^{(1)}$, A. J. Millis$^{(2)}$, and S. Das Sarma$^{(3)}$}

\address{
(1) Physics Department, Harvard University, Cambridge, MA 02138
\\
(2) Department of Physics, Colombia University, New York, NY 10027
\\
(3) Condensed Matter Theory Center, Department of Physics,
University of Maryland,
College Park, MD 20742
}

\date{\today}

\begin{abstract}
In this paper, we review recent development in the theory of resonant 
inelastic light (Raman) scattering in one-dimensional electron systems.
The particular systems we have in mind are electron doped 
GaAs based semiconductor quantum wire 
nanostructures, although the theory
can be easily modified to apply to other one-dimensional systems. 
We compare the traditional conduction-band-based non-resonant
theories with the full resonant theories including the effects of interband 
transitions. We find that resonance 
is essential in explaining the experimental data in which the single
particle excitations have finite spectral weights comparable to the 
collective charge density excitations.
Using several different theoretical models (Fermi liquid model, 
Luttinger liquid model,
and Hubbard model) and reasonable approximations, we further
demonstrate that the ubiquitously observed strong single
particle excitations in the experimental Raman spectra 
cannot be explained by the spinless multi-spinon 
excitations in the Luttinger liquid description. The 
observability of distinct Luttinger
liquid features in the Raman scattering spectroscopy is critically
discussed.
\end{abstract}


\maketitle

\section{Introduction}
\label{introduction}

One-dimensional (1D) electron systems, where electron dynamics is 
constrained to be along a single direction
(chosen as the $x$ axis in the rest of this paper where necessary)
due to the quantum mechanical confinement of the 
carrier system imposed by suitable externally
applied electrostatic potentials along $y$ and $z$ directions
(leaving the $x$ direction to be the only direction of
free-electron-like motion characterized by a 1D wavevector
$k$), have been achieved in the electron-doped GaAs quantum wire
structures by combining the state of the arts semiconductor materials
growth with extremely clever nanolithographic fabrication technique 
\cite{technique}.
In these 1D semiconductor quantum wire structures
the noninteracting 3D electron wavefunction can be described, to 
a very good level of accuracy, using
the effective mass approximation \cite{Lai,effective_mass2,effective_mass3} as 
$\Psi(\bfr=(x,y,z))=\tilde{\Psi}_j(y,z)
\,e^{ikx}/\sqrt{L_x}$, where $L_x$ is the plane wave
normalization length along the wire direction, $x$, and 
$\tilde{\Psi}_j(y,z)$ is the bound wavefunction for electron motion in 
the quantized transverse ($y-z$) direction with 
$j$ denoting a particular bound state \cite{Lai} for the 
$y-z$ motion. The transverse bound states (usually called
"subbands" in the semiconductor literature \cite{Lai,dassarma_review}) 
characterized by
the discrete index $j$ ($=0,1,2,\cdots$ with '$0$' being
the ground state lying lowest in energy near the conduction
band minimum of GaAs) are typically
separated by a few meV in energy with their
separation (as well as the carrier density in the system)
being somewhat tunable through various gate voltages 
applied from outside. For low temperature ($\leq 1$ K),
$k_B T\ll (E_1-E_0)$ where $E_j$ is the $j$th subband
energy for transverse motion and therefore $(E_1-E_0)$,
the lowest intersubband energy separation, is the low-lying
excited state energy, the quantum wire system is by definition,
a strictly one-dimensional quantum mechanical electron
system at low carrier densities [i.e. for $E_F<(E_1-E_0)$] so
that only the lowest quantum level, the ground subband,
is occupied by electrons. Even in a situation where
$E_F>(E_j-E_0)$ for a few values of $j$, the semiconductor 
quantum wire system is a "multisubband" 1D electron 
system \cite{dassarma_review}
as long as the intersubband scattering between different subbands is 
relatively weak (which is usually the case).

Such 1D semiconductor quantum wires, particularly in their strict 1D 
one-subband [i.e. $E_F<E_1-E_0$] limit, are examples of interacting
1D electron systems, the so-called Tomonaga-Luttinger liquids
("Luttinger liquids"), which are of great intrinsic and fundamental interest 
in condensed matter physics \cite{ll,voit95,schulzemery,manhan}. 
In particular, Luttinger liquids (LL)
are fundamentally different from Fermi liquids (i.e. interacting 2D and 3D
electron systems such as normal metals and two-dimensional electron systems
confined in semiconductor heterostructures) in the sense that
the one-to-one correspondence between the interacting ("Fermi liquids")
and the noninteracting ("Fermi gas") systems, which is the basis of the 
very successful Landau Fermi liquid (FL) theory in two- and three-dimensional 
electron systems, categorically breaks down for 1D Luttinger liquids.
As a result, one-dimensional electron systems do not have a Fermi
surface defined by a finite jump of momentum distribution, $n_{\bf k}$, 
at Fermi wavevector at zero temperature even in the presence of 
weak interaction
(more precisely, if one defines the Fermi surface to be a singularity
of $n_{\bf k}$ at $|{\bf k}|=k_F$, then 1D interacting electron systems 
can still have a Fermi surface due to the infinite slope of 
$n_k$ at $|k|=k_F$), i.e.
interaction effects are nonperturbative in one-dimensional electron
systems.
%
%

Luttinger liquids (i.e. interacting 1D electron systems) are characterized by 
the absence of long wavelength low energy single particle (i.e. electron-hole) 
excitations which dominate the low energy spectra of 2D and 3D systems 
and by the existence of spin-charge separation, i.e. separate branches of 
low energy excitations in Luttinger liquids can carry spin but no charge 
("spinons") or can carry charge but no spin ("holons")
in contrast to higher dimensional systems where the single-particle
excitations necessarily carry both spin and charge. The zero-temperature
momentum distribution function in Luttinger liquids does not have the usual 
discontinuity at $k=k_F$ indicating the existence of a Fermi surface, but 
instead has a power-law behavior $n(k\sim k_F)\sim 
\frac{1}{2}-{\rm sgn}(k-k_F)|k-k_F|^\alpha$,
where the exponent (the so-called Luttinger exponent) is non-universal.

In this paper we review our recent theoretical work on the inelastic 
resonant light ("Raman") scattering studies of 1D semiconductor quantum 
wires, using both the Fermi liquid and 
the Luttinger liquid approaches. Resonant
Raman scattering (RRS) has been a very successful tool for studying the
elementary electronic excitation spectra in doped semiconductors.
In particular, the mode dispersion 
(i.e. the frequency as a function of wavevector)
and the spectral weight (i.e. the oscillator strength) of low energy (from
a fraction of an meV to tens of meV) electronic excitations can be directly 
obtained via resonant Raman scattering (in the $10^5-10^6$ cm$^{-1}$ 
wavevector range). Since the interesting and important elementary electronic
excitations in GaAs quantum wire (1D) and quantum well (2D) structures
lie precisely in this frequency-wavevector range, resonant Raman scattering 
spectroscopy has been an effective tool for studying electronic
excitation spectra in GaAs based low dimensional electron systems
over the last twenty-five years 
\cite{dassarma_review,pinczuk,pinczuk_1D,pinczuk_2D,rrs_exp96}. 
In addition, various selection rules
involving the relative polarization of the 
incident and scattered photons in the 
Raman spectra (the so-called polarized or the depolarized spectra)
allow one to study charge density or spin density
excitations in the system, making the resonant Raman scattering 
spectroscopy a rather powerful tool for studying intra- and inter-subband 
electronic excitations in 1D and 2D electronic systems including
the strongly correlated fractional quantum Hall regime \cite{QH_exp}.

The Raman scattering spectra, within the simple linear response 
theory, is directly proportional to the dynamical structure factor
of the interacting electron system which, at long wavelength,
has significant spectral weight only at collective excitations, if
no inter-subband resonant scattering involved 
\cite{dassarma_review,pines,3D,2D}. Depending
on the polarization configuration of the experimental set up, the 
Raman scattering spectra should directly measure either the 
collective charge density excitation (in the polarized or the non-spin-flip
configuration) or the collective spin density excitation (in the
depolarized or the spin-flip configuration) 
\cite{pinczuk,pinczuk_1D,QWR_ref,rrs_exp96}. 
Within the simple linear
response theory \cite{dassarma_review,pines,3D,2D}, 
the Raman scattering spectra in the two 
configurations is simply proportional to the
imaginary part of the screened (for the charge density excitation
(CDE) in the polarized configuration) or unscreened 
(for the spin density excitation (SDE) in the depolarized channel)
polarizability function.
At very small wavevectors that can be probed in the Raman 
scattering experiments, only the collective modes should have 
appreciable spectral weight in the Raman scattering spectra.
This fact, i.e. that only collective modes (either CDE in the
polarized spectra or SDE in the depolarized spectra) can manifest 
themselves in the inelastic light scattering spectra of semiconductor
structures applies to systems of any dimensionality, 3D, 2D or 1D 
electron systems \cite{dassarma_review,3D,2D,wang02_nrs}. 
Of course, in principle, the incoherent single particle
excitations (electron-hole pairs) can be also observed in the Raman
scattering experiment of two- and three- dimensional system with small
(but finite) momentum tranfer, due to the existence of quasi-particle
excitations. In one-dimensional electron system, on the other hand,
only collective charge and spin modes are expected to be observed 
since the single particle excitations are absence in the
LL theory as we mentioned above.
%
%

A real intriguing aspect of the resonant inelastic light scattering 
spectroscopy in semiconductor structures has, however, been the 
persistent and ubiquitous presence of a "single-particle excitation"
(SPE) peak in the Raman scattering spectra in sharp contradiction
with the simple theoretical description provided above. This single
particle excitation peak, which is usually fairly weak (but orders 
of magnitude stronger than that given by the simple electronic 
response function argument given above), is almost always present
in the resonant Raman scattering spectra in addition to the expected
peaks associated with the collective excitations 
\cite{pinczuk_1D,pinczuk_2D,QWR_ref,rrs_exp96,Jusserand00,new_data}.
A very interesting aspect of this phenomenon is the fact that
collective mode spectral features in the resonant Raman 
scattering spectra seem to be rather well-described by the 
simple response theory, which at the same time predicts orders of
magnitude weaker values for the single particle spectral weight
(in 2D and 3D) than that observed experimentally. In 1D semiconductor
quantum wire structures, where the Luttinger liquid behavior
manifestly precludes the existence of low-lying single particle
excitations, the observed existence of SPE features in the 
Raman scattering spectra \cite{pinczuk_1D,QWR_ref,new_data} 
raises very serious conceptual
questions regarding our basic understanding of the elementary
excitation spectra in 1D electron systems.

\begin{figure}
 \vbox to 5cm {\vss\hbox to 5cm
 {\hss\
   {\includegraphics{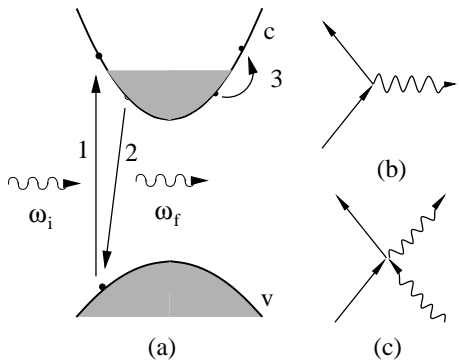}
   }
  \hss}
 }
\caption{(a) Schematic representation of the two-step RRS in the 
direct gap two band [c(v): conduction (valence) band] 
model. $\omega_i$ and 
$\omega_f$ are the initial and final frequencies of the external photons. 
(b) and (c) are the
Feynman diagrams of the electron-photon scattering process described by 
$\bfp\cdot\bfA$ and $\bfA\cdot\bfA$ terms respectively in the interacting 
Hamiltonian (see text). Solid and wavy lines represent the electron and 
photon Green's functions respectively.
}
\label{RRS}
\end{figure}

In this review we discuss how this conceptual problem has 
recently been resolved theoretically by showing that a full 
understanding of the existence of single particle excitation like
spectral features in the {\it resonant} Raman scattering spectra 
necessarily requires going beyond the simple {\it non-resonant}
single band (i.e. conduction band) linear response theory and 
considering the full "two-step" {\it resonant} aspect of the experiment
(Fig. \ref{RRS}(a)) where the valence band plays a crucial 
role \cite{Sassetti98}. 
For completeness we will first review the theories of 
nonresonant Raman scattering using 1D FL model, LL model, 
and lattice Hubbard model respectively. We then
discuss the theories of resonant Raman scattering including
the two-step scattering process. Finally, we discuss how these results
are related to the existing experimental data and their implication
to the Luttinger liquid properties in 1D electron-doped semiconductor
quantum wire systems.

In Fig. \ref{RRS}(a) we depict the schematic 
diagram \cite{Dassarma99} for the
two steps involved in the
resonant Raman scattering process: an electron in the valence
band is excited by the incident photon into the conduction 
band above Fermi surface, 
leaving a valence band hole behind (step 1), and
then an electron from inside the conduction band Fermi surface recombines with
the hole in the valence band (step 2), emitting an outgoing photon with 
an energy and momentum (Stokes) shift.
(In principle, these two steps could occur in different orders.) 
The net result is an elementary electronic 
excitation created in the conduction band through the 
intermediate valence band states. The 
{\it non-resonant} approximation to RRS ignores the intermediate 
valence band states as shown by the step 3 in Fig. \ref{RRS}(a). 
Note that the resonant process
depends on the incident photon energy, while
the non-resonant approximation depends 
only on the energy difference between the incident and the
scattered photons. This difference
turns out to be crucial in the RRS theory as shown below.
Total electron spin is conserved in the final scattering processes since
we are considering only the polarized geometry (i.e. photon polarization
is not changed).
We restrict ourselves to the non-spin-flip polarized RRS, 
where the CDE dominates the non-resonant linear response 
spectra.

\section{Nonresonant Raman scattering theory}
\label{NRS_theory}

In the presence of an external photon field 
the interaction between
the electron gas and the radiation field is described by 
the following Hamiltonian:
\begin{eqnarray}
H&=&H_{e}+\sum_i^N\left[-\frac{e}{m_i c}\bfp_i\cdot\bfA_i
+\frac{e^{2}}{2 m_i c^{2}}
\bfA_i^{2}\right],
\label{H_tot}
\end{eqnarray}
where $\bfA_i=\bfA(\bfx_i,t)$ is the vector potential of photon.
$\bfx_i$ and $\bfp_i$ are the position and momentum operators 
of $i^{\rm th}$ electrons, and
$c$ is the speed of light. $m_i$ is the 
effective electron mass in the semiconductor bands
($m_i$ can be different if considering interband scattering). 
$H_e$ is the Hamiltonian of electrons
interacting with Coulomb potential without the radiation field. 
Figs. \ref{RRS}(b) and 
\ref{RRS}(c) correspond to the scattering processes induced by
the linear ($\bfp\cdot\bfA$) term and the quadratic ($\bfA^{\rm 2}$) 
term respectively in the second quantization representation. 
One can simply neglect the $\bfp\cdot\bfA$ term in Eq. (\ref{H_tot}) if
only the $non$resonant Raman scattering spectroscopy is considered,
where the incident photon frequency is far away from the band gap energy
\cite{Walf,sakurai}.
The resulting Raman scattering intensity therefore is equivalent
to the imaginary part of the time-ordered density correlation
function in the linear response theory~\cite{pines,fetter}:
\begin{eqnarray}
&&{\rm Im}\left[i\int_{-\infty}^\infty dt e^{i\omega t}
\langle T[ n^\dagger(k,t)n(k,\rm{0})]\rangle_{\rm{0}}\right],
\label{Pi}
\end{eqnarray}
where $\langle\cdot\cdot\cdot\rangle_0$ is the ground 
state expectation value, and 
$n(k,t)$ is the electron density operator.
In the rest of the this section, we will compare the results of 
Eq. (\ref{Pi}) calculated by different theoretical models.

\subsection{Fermi liquid model}
\label{NRS_FL}

In the Fermi liquid model \cite{pines}, the elementary excitations of an
interacting electronic system are quasi-particles,
which have the same quantum numbers as free electrons but
with an effective mass and renormalized single particle parameter.
It is well-known that the Fermi liquid model is a very good 
approximation in two and three dimensional systems, but fails in 
one dimensional systems due to strong fluctuations.
However, it has also been noticed that \cite{wang02_nrs,larkin}
the collective plasmon modes
calculated by the standard random phase approximation (RPA)
within the FL model is exactly the same as the one obtained by
Luttinger liquid model (see below). Its energy dispersion is 
$\omega_\rho(k)=v_F|k|\sqrt{1+2V_c(k)/\pi v_F}$, where $v_F$
is Fermi velocity and $V_c(k)\propto \ln (1/kd)$ is the 1D
Coulomb interaction with $d$ being the characteristic confinement 
length in the transverse dimension. Therefore it is instructive
to compare the RRS spectrum calculated in the FL model within RPA
to the results obtained by other exactly solvable models in
1D electron system (see below).

In Fig. \ref{spectrum_rpa}, we show
the typical nonresonant Raman scattering spectra calculated in the FL model
within RPA (solid lines). It shows a strong
CDE spectral weight at the plasmon mode energy, and a
much weaker (three orders in magnitude) weight in the single
particle excitation (SPE) energy, $\omega=kv_F$. The almost vanishing weak
presence of the SPE peak results from the quasi-particle excitations in the FL 
model. The dashed lines in the same figure are results calculated 
by including vertex corrections  
(within Hubbard approximation \cite{wang02_nrs,hubbard}) 
in the theory to go beyond the RPA approximation. 
We find that the simple non-resonant vertex correction still 
gives qualitatively the 
same result with a SPE spectral weight orders of magnitude 
weaker than the CDE.
Including the effects of nonparabolicity of the electron band energy
and/or the effects of the breakdown of electron momentum by scattering
with impurity potential can enhance the SPE weight slightly (less than one 
order of magnitude) but 
still does not change the picture qualitatively \cite{wang02_nrs}.
Thus, possible adjustments and improvements of the
theory staying within the conduction band nonresonant Raman scattering
picture are not capable of explaining the experimental observation
of a strong presence of the SPE spectral feature in the RRS 
spectra \cite{QWR_ref}.
\begin{figure}
 \vbox to 5cm {\vss\hbox to 5.5cm
 {\hss\
   {\includegraphics{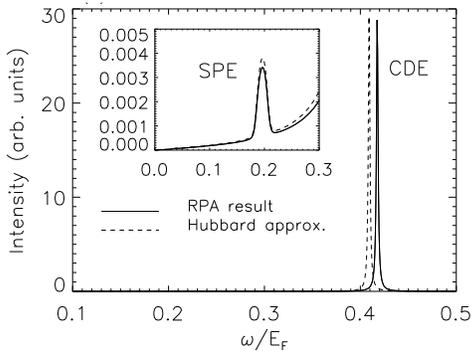}
   }
  \hss}
 }
\caption{
Dynamical structure factor of a 1D electronic systems 
obtained by the standard 
(nonresonant) RPA calculation at $k=0.1 k_F$. The electron densities
used in the calculation is $6.5\times 10^5$ cm$^{-1}$. Finite impurity
scattering ($\gamma=10^{-3}E_F$) has been applied to broaden the peaks.
}
\label{spectrum_rpa}
\end{figure}

\subsection{Luttinger liquid model}
\label{NRS_LL}

The Luttinger liquid model~\cite{ll,voit95,schulzemery,manhan}
is thought to provide a generic low energy description for 1D electron
systems, which are characterized by the LL
fixed point in the renormalization group sense. The standard and exactly
solvable LL model is the 1D electron gas with a linear dispersion 
($E_k=rv_F(k-rk_F)$) around Fermi points ($\pm k_F$) at each branch 
($r=\pm 1$) and with short-ranged forward interaction~\cite{ll,voit95}. 
It is well-known \cite{voit95} that the exactly diagonalized LL Hamiltonian 
consists of two independent elementary excitations:
charge bosons (holons) and spin bosons (spinons), 
the so-called spin-charge separation.
The former is essentially equivalent to the spinless 
charge density excitations of the FL model with the same plasmon velocity, 
while the latter occurs
in the depolarized (spin-flip) scattering channel at the Fermi velocity,
and is akin to the spin density excitation mode of the FL.
Therefore, in the nonresonant polarized Raman scattering spectroscopy
we consider in this paper, the LL model has only the charge boson (plasmon)
excitations, and does not have any single particle
weight due to the breakdown of Landau Fermi liquid in 1D 
system. However, including the nonlinearity of the band energy
may lead to situation where the
charge mode and the spin mode sectors interact with each other and
cause possible multiboson excitation above the Fermi surface.
It has been proposed \cite{schulz93} 
that a spin singlet excitation (SSE) of two bound spinons
(of total spin zero) may be responsible for the observed single particle 
excitation in the Raman scattering experiments.
It is therefore  important to study how such multi-boson
excitations affect the polarized spectrum in an exactly solvable 
model with a nonlinear band energy. We therefore consider the 1D Hubbard 
model in the following subsection in this context, considering
in details its excitation spectra. Although the lattice Hubbard model
does not really apply to continuum semiconductor quantum wire 
systems, generic LL properties (e.g. the excitation spectra and 
spectral weights) should be independent of the model.

\subsection{Hubbard model}
\label{NRS_HM}

The exactly solvable 1D single band
Hubbard model (HM) contains a hopping matrix element between
neighboring sites, $t$, and a spin-{\it dependent} on-site
interaction, $U$. The full Hamiltonian is 
\begin{equation}
H=-t\,\sum_{i,\sigma}\left(c^{\dagger}_{i+1,\sigma}c_{i,\sigma}+
\mathrm{H.c.}\right)+U\sum_{i}n_{i\uparrow}n_{i\downarrow},
\label{HM_eq}
\end{equation}
where $c_{i,\sigma}$ and $n_{i,\sigma}$ are respectively
the fermion creation operator and the density
operator for site $i$ and spin $\sigma$.
Among the many accurate and useful methods to study the 1D HM,
we use the Bethe-ansatz method~\cite{lieb,coll,yang}
to obtain the ground state and the low-lying excitation spectra. 
Since the Bethe-ansatz
wavefunctions are not particularly useful in calculating the
correlation functions,
we use the Lanczos-Gagliano (LG) diagonalization method
\cite{wang02_nrs,gagliono} to directly calculate the spectral weights
of these elementary excitations.
Our results obtained by this technique are consistent 
with the quantum Monte Carlo
calculations~\cite{num_sc_separation} where appropriate.

\begin{figure}

 \vbox to 5cm {\vss\hbox to 6cm
 {\hss\
   {\includegraphics{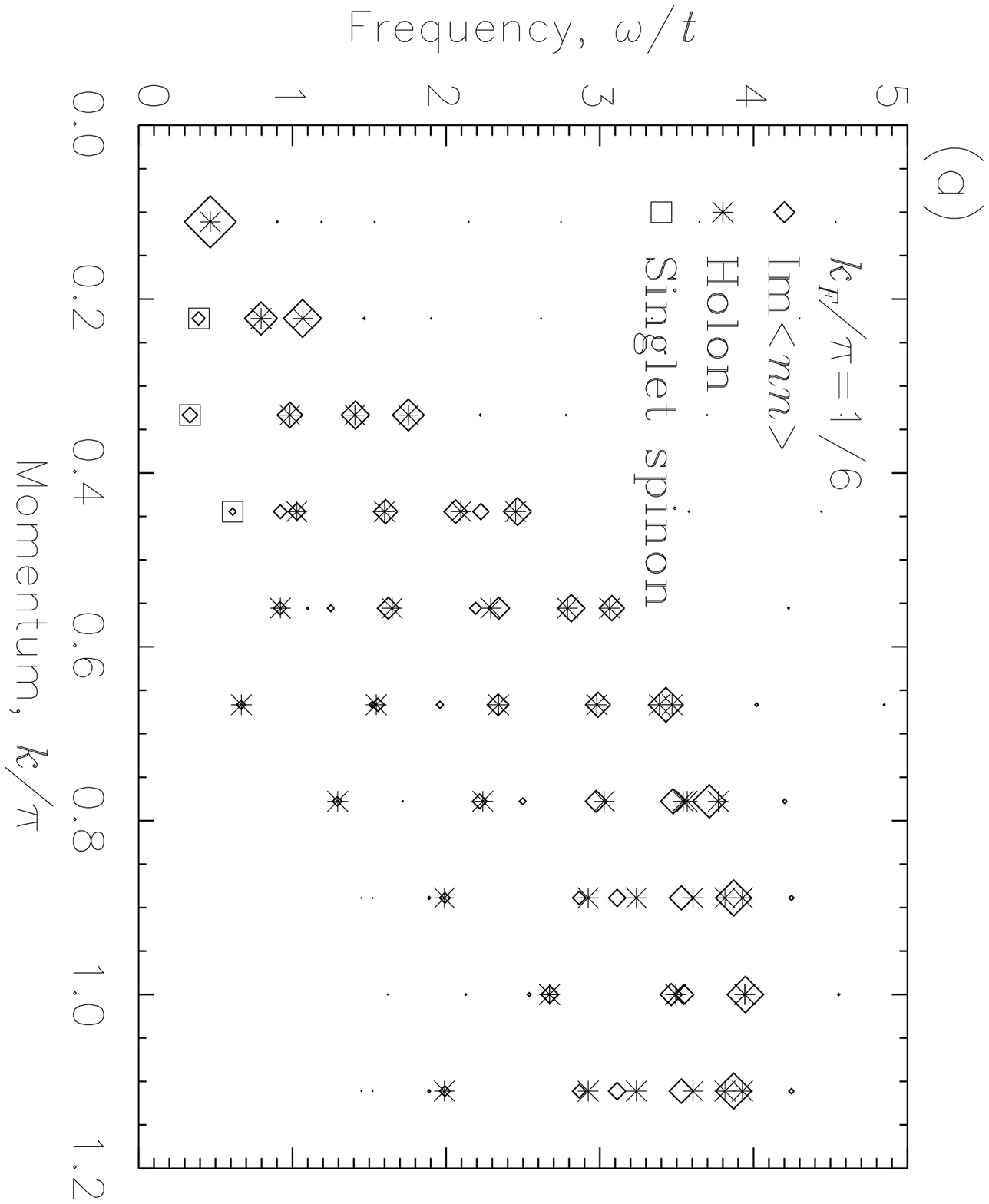}
   }
  \hss}
 }
 \vbox to 5cm {\vss\hbox to 6cm
 {\hss\
   {\includegraphics{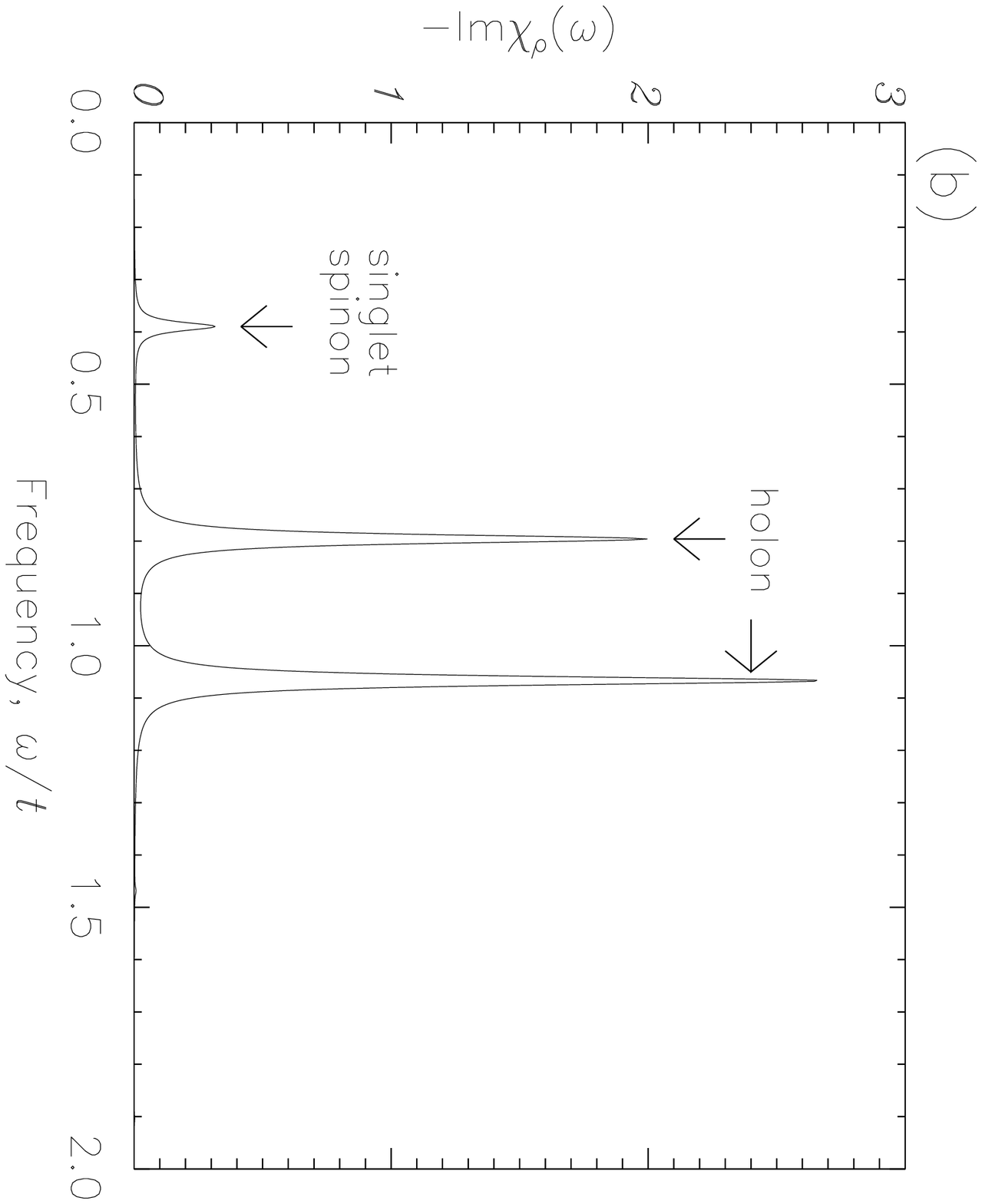}
   }
  \hss}
 }
\caption{
(a) Energy-momentum dispersion and (b) the
spectrum of charge density correlation function 
of 1D HM for 6 electrons in 18 sites with $U/t=3$.
$k=2\pi/9$ for the spectrum (b). Holon excitations here are 
equivalent to the charge density excitations in the Raman 
scattering experiments.
The area of each diamond(square) in (a) is proportional to the
spectral weight of each charge(spin) excitation peak. 
}
\label{HM}
\end{figure}

In Fig. \ref{HM}(a), we show the energy-momentum dispersion obtained 
from the poles of the imaginary part of the charge density correlation
function together with the results calculated by Bethe-ansatz equations.
The center of each open diamond represents the position of the pole, and
its area is proportional to the spectral weight of that excitation.
We find that the charge density excitations (often these 
excitations are called holons or particle-hole excitations in the 
Bethe-ansatz literature \cite{schulzemery,coll})
cover almost exactly the same region
including the energy minimum at $4k_F$ except for the lower-lying peaks
around $2k_F$, where the singlet spinon just matches those peaks.
In Fig. \ref{HM}(b), 
we show the imaginary part of the charge density correlation
function of the same system at $k=2\pi/9$. It shows
that singlet spinons have a relatively small but non-negligible weight
(different from the results of FL and LL models),
compared with the weight of the dominant charge density excitations (holons).
Their relative spectral weight ratio is less than 0.1.
We have also studied the dispersions and spectral weights
of different filling factors and/or different interaction strengths
(for more details, see Ref. [\onlinecite{wang02_nrs}]), but 
do not see any possibility to obtain a reasonable fit of
the "two peak" RRS structure observed in Ref. [\onlinecite{QWR_ref}].
Therefore, we conclude that although the nonparabolicity of 
the electron conduction band and the spin-dependent interaction
contained in the Hubbard model enhance the spectral weight 
of the singlet spinon excitations (which can be interpreted as the 
SPE feature in the RRS experiments), the present results obtained without
resonance effects cannot explain the experimental data 
in the RRS experiments. The obtained SPE-like RRS feature 
cannot be  explained by staying within a single band LL model.

\section{Resonant Raman scattering theory}
\label{RRS_theory}

We now consider the full resonance situation (step 1 and 2 in Fig. \ref{RRS})
of a Raman scattering process by including the valence band 
explicitly \cite{Dassarma99,Jusserand00,Sassetti98,Wang00}. 
When the incident photon energy is near the
$E_0+\Delta_0$ direct gap,
the second order perturbative contribution of
the $\bfp\cdot\bfA$ term in Eq. (\ref{H_tot}) becomes important and comparable
to the first order contribution of the $\bfA^{\rm 2}$ term,
leading to an electron
interband transition between the conduction band and the valence band.
The finite time duration between the first step and the
second step of the scattering process 
gives a nontrivial contribution to the scattering matrix element.
The transition rate in the second order
perturbation theory can be obtained to be \cite{wang02_rrs}
(we assume the electron-photon coupling vertex to 
be a constant for simplicity)
\begin{widetext}
\begin{eqnarray}
W&=&\lim_{T\rightarrow\infty}\frac{1}{T}
\sum_{\bfp_{\rm 1},\bfp_2,\sigma_1,\sigma_2}
\int_{-T/2}^{T/2}dt_{1}\int_{-T/2}^{t_1}dt_{2}\int_{-T/2}^{T/2}
dt'_1\int_{-T/2}^{t'_1}dt'_2
\,e^{i\bar{\omega}(t'_2-t'_1+t_1-t_2)}e^{i\omega(t_2'+t_1'-t_1-t_2)/2} 
\nonumber\\
&&e^{iE^v_{\bfp_1}(\it{t'}_{\rm 1}-\it{t'}_{\rm 2})}
e^{iE^v_{\bfp_3}(\it{t}_{\rm 1}-\it{t}_{\rm 2})}
\langle c_{\bfp_1+\bfq/2,\sigma_{\rm 1}}(t_2')
c^{\dagger}_{\bfp_{\rm 1}-\bfq/2,\sigma_1}(t_1')
c_{\bfp_{\rm 2}-\bfq/\rm{2},\sigma_2}(t_1)
c^{\dagger}_{\bfp_{\rm 2}+\bfq/2,\sigma_{\rm 2}}(t_2)\rangle_0,
\label{W}
\end{eqnarray}
\end{widetext}
where we have chosen the backward scattering channel
$\bfk_i=-\bfk_f=\bfq/\rm{2}$ 
and $\omega_{i,f}=\bar{\omega}\pm\omega/2$, without any loss of generality. 
$E^v_\bfp$ is the band energy of electrons
in the valence band. This result can be evaluated within the Fermi 
liquid model and the Luttinger liquid model independently and we present
these results respectively in the following
sections. In Eq. (\ref{W}) and the following formula in the FL 
model, we keep the vector form of the momentum indices because they
apply equally well to two- and three-dimensional systems.

\subsection{Fermi liquid model}
\label{RRS_FL}

As mentioned above, in the FL model, the single particle energy
is assumed to be well-defined, so that one can easily integrate out 
the time difference between the absorption and the emission of 
the external photon and obtain \cite{wang02_rrs}
\begin{eqnarray}
W(q,\omega;\Omega)&=&\int_{-\infty}^\infty dt\,e^{i\omega t}\langle 
N^\dagger(\bfq,t)N(\bfq,\rm{0})\rangle_{\rm 0},
\label{W_rrs_rpa}
\end{eqnarray}
where the resonant "density" operator, $N(\bfq,t)$, is defined to be
$N(\bfq,t)
=\sum_{\bfp,\sigma}A(\bfp,\bfq)c^\dagger_{\bfp+\bfq/\rm{2},\sigma}(t)
c_{\bfp-\bfq/\rm{2},\sigma}(t)$
with the matrix element $A(\bfp,\bfq)$:
\begin{eqnarray}
A(\bfp,\bfq)&=&\frac{1}{-{\Omega}+(1+\xi)(E^c_\bfp-E_F)
+E^c_\bfq/\rm{4}+\it{i}\lambda}.
\label{A}
\end{eqnarray}
Here ${\Omega}\equiv \bar{\omega}-E_g-(1+\xi)
E_{F}$ is the mean photon energy relative to the resonance energy,
and $\xi\equiv m_c/m_v$ is the ratio of the carrier effective mass
in conduction and valence bands; $E_F=E^c_{k_F}=k_F^2/2m_c$ 
is the Fermi energy of the conduction band electrons.
$\lambda$ is a phenomenological broadening parameter we introduce to include
roughly all possible broadening effects during the 
resonance scattering process.
(A microscopic evaluation of $\lambda$ seems to be essentially
impossible at the present time. \cite{endnote0})

Comparing Eq. (\ref{Pi}) with Eq. (\ref{W_rrs_rpa}), we 
find that the resonance effect
on the conduction band electrons is in the matrix element 
$A(\bfp,\bfq)$, which arises from the time difference between the
two steps of Raman scattering.
In the following discussion we define "off resonance" 
as $|\tilde{\Omega}|> E_F$ and "near resonance" as 
$|\tilde{\omega}| \ll E_F$. 
Off resonance the spectral weight decreases as $|\tilde{\Omega}|^{-2}$,
while near resonance the singular properties of 
$A(\bfp,\bfq)$ strongly enhance the spectral weight nontrivially.  
The calculation of the RRS spectrum is therefore reduced to the
evaluation of the correlation function of
Eq. (\ref{W_rrs_rpa}), which can be easily calculated within 
the RPA approximation \cite{wang02_rrs}.

\begin{figure}
 \vbox to 8.5cm {\vss\hbox to 6cm
 {\hss\
   {\includegraphics{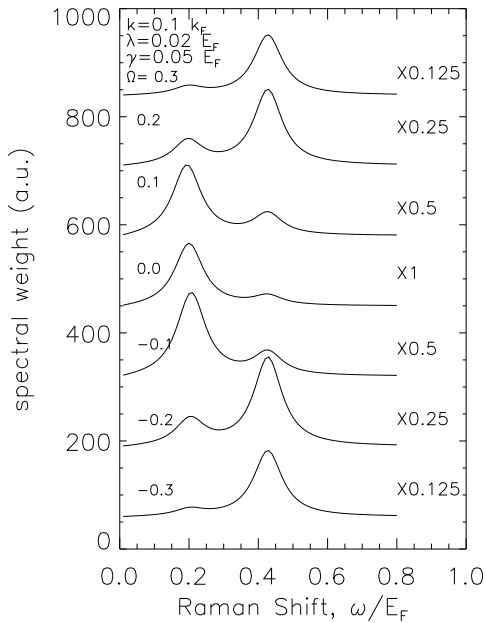}
   }
  \hss}
 }
\caption{Raman scattering spectrum near resonance calculated in the FL model.
Finite impurity scattering ($\gamma$) 
has been included to broaden the resonance peak 
properly. Other system parameters are the same
as used in Fig. \ref{spectrum_rpa}. Note that the scales of each plot
are indicated in the right hand sides.
}
\label{spectrum_rrs_rpa}
\end{figure}
In Fig. \ref{spectrum_rrs_rpa}, we show a typical result of the 
resonance Raman scattering spectra in the polarized
channel. We find that the resonance effects strongly enhance the SPE
spectral weight near resonance ($|{\Omega}|\leq 0.1 E_F$), 
making the SPE weight even larger than the CDE spectral weight.
Off resonance ($|\tilde{\Omega}|>0.1 E_F$), the SPE weights 
become much smaller than the CDE weight very similar to
the nonresonance situation. We note that the Raman energy
shift is not affected by resonance effects and hence the one-band 
nonresonant linear response theory still well-describes the 
spectral peak energy dispersions. 
This agrees well with the existing RRS
experimental date \cite{QWR_ref,rrs_exp96,new_data}.

\subsection{Luttinger liquid model}
\label{RRS_LL}

The general formula of Raman scattering spectra, Eq. (\ref{W}), can also be
evaluated within the LL model.
Using the space-time translational symmetry, 
Eq. (\ref{W}) can be simplified by representing 
the fermion operators in the coordinate space:
\begin{equation}
W(q,\omega;\Omega)=\int dRdTe^{i(\omega T-qR)}\langle
\widehat{O}^{\dagger}(R,T)\widehat{O}(0,0)\rangle_0,
\label{W_O}
\end{equation}
where
\begin{eqnarray}
&&\hspace{-0.5cm}\widehat{O}(R,T)=\sum_{r,s}
\int dx\int_{0}^{\infty}dt\,\phi(x,t)
\nonumber\\
&&\times \psi_{r,s}(R+x/2,T+t/2)\psi _{r,s}^{\dagger}(R-x/2,T-t/2).
\label{O}
\end{eqnarray}
$\psi_{rs}$ is the electron operator for the left ($r=-1$) and the 
right ($r=+1$) fermion branch of spin index $s$. The retardation function, 
$\phi(x,t)$ is
\begin{eqnarray}
\phi (x,t) &=& \frac{e^{i\bar{\omega} t}}{L}\sum_{p}e^{i(E_{p}^{v}t-px)}
\nonumber\\
&=&e^{i({\Omega}+v_{F}^{v}k_{F})t}\delta(x+rv_{F}^{v}t),
\label{phi}
\end{eqnarray}
where we have used the linearized valence band energy around
the Fermi wavevector, and $v_F^v$ is the associated valence band velocity.
Eqs. (\ref{O})-(\ref{phi}) are the fundamental results of the RRS theory 
in the LL model. They show that the RRS
process creates an electron-hole pair separated in space by 
$x$ and in time by $t$, with the amplitude for a given space-time
separation controlled by the function $\phi(x,t)$.
Far from resonance, $\phi(x,t)$ is short
ranged in both $x$ and $t$, so that $\widehat{O}$ becomes similar to 
the ordinary
density operator and $W$ becomes the charge density correlation function
\cite{Wang00,wang02_rrs}(e.g. Eq. (\ref{Pi})). As the
mean photon energy is tuned closer to the resonance condition, 
$\phi(x,t)$ becomes longer
ranged, and then $\widehat{O}$ becomes nonlocal in both space and time.
Similar to the FL model, this non locality will be seen to give rise to 
the interesting resonance effects, by allowing the light to couple
to something other than the dynamical structure factor, making the situation
qualitatively different from the nonresonant one-band situation.

\begin{figure}

 \vbox to 8.5cm {\vss\hbox to 5.cm
 {\hss\
   {\includegraphics{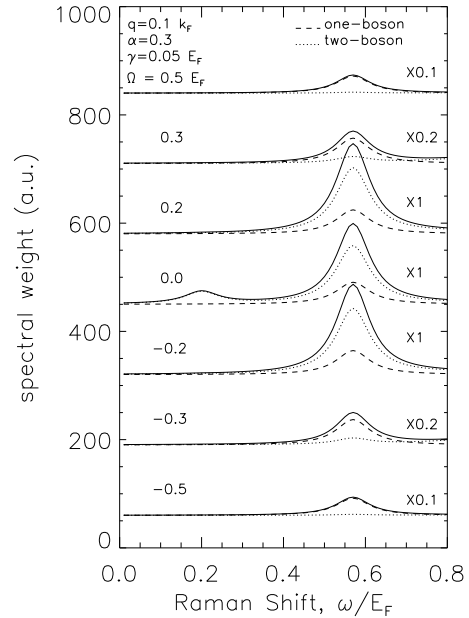}
   }
  \hss}
 }
\caption{
Calculated polarized RRS spectra for various resonance condition,
$\tilde\Omega$, in LL model.
One- and two-boson contributions have been plotted separately in order to show
their relative contributions. The Fermi velocity is the same
as used in Fig. \ref{spectrum_rrs_rpa}, while the short-ranged 
interaction strength is chosen to $\alpha=0.3$.
}
\label{12_rrs}
\end{figure}

Although in principle Eq. (\ref{W_O}) can be reduced further by using
the bosonization method and then calculated numerically,
it is more instructive to consider their leading order contributions from
the one-boson and two-boson excitations \cite{Wang00}.
The former is directly related to the usual plasmon mode excitation
(i.e. CDE) in the FL-RPA theory, and the
latter is associated with the singlet two-spinon excitations 
at $\omega=q v_F$. The analytical results for these two leading order
contributions can be also obtained \cite{Wang00,future}, and one finds 
that the spectral weight of the charge boson mode (i.e. CDE)
decreases as $|\Omega|^{2\alpha-2}$ at off resonance \cite{Wang00,endnote}  
where $\alpha\in [0,1)$ is the Luttinger liquid exponent 
which is positive for repulsive interaction. 
In Fig. \ref{12_rrs}, we show the calculated polarized LL RRS spectra
including one and two boson contributions for different resonance conditions.
One observes that near resonance the "SPE" peak
(now it is composed by two spinon excitations) is noticeable, but 
still has rather weak spectral weight compared with the CDE 
(charge boson) peak.
This is because any resonance enhanced single particle
excitation during the RRS process will be immediately 
separated into spin and charge
channels in the LL model due to the spin-charge separation.
The angular momentum conservation imposed selection rules for 
non-spin-flip scattering processes automatically suppress 
the single spin-boson contribution, so that only the singlet
spinon excitations (composed by at least two spinons)
can contribute to the spectral weight. Therefore, the strongly 
suppressed "SPE" mode in the polarized
spectrum near resonance may possibly be a characteristic LL
signature in the semiconductor quantum wire systems.

We also have studied the transition rate, 
Eq. (\ref{W_O}), by summing all higher
order results beyond
the leading one and two boson contributions. We find that
\cite{future} the spectral weight of the 
total charge boson excitation (i.e. CDE) is
still much larger than the total singlet spinon 
excitation (SSE) in the parameter
regime of the existing experiments. The ratio (CDE/SSE) becomes
close to unity only near the noninteracting limit, but increases 
when the electron-electron interaction becomes stronger. We have also 
considered the situation of the long-ranged Coulomb interaction rather
than the short-ranged interaction extensively used in the standard LL model. 
We find that the SSE spectral weight is further suppressed by the 
long-ranged Coulomb 
interaction, showing that the weak singlet spinon excitations in the
polarized RRS spectroscopy are generic features of the LL model.
Therefore we conclude that the experimental RRS results obtained 
so far in the literature \cite{pinczuk_1D,QWR_ref,new_data} 
are not decisive signatures of  
Luttinger liquids in the semiconductor quantum wires since
the SSE spectral weight seems to remain somewhat weak 
in the LL theory compared with the RRS observations (and
the SSE is really the only available candidate for
the SPE seen in the RRS spectra within the LL model).

\section{Discussion}
\label{discussion}

As we have mentioned briefly in the context of Eq. (\ref{phi}), the most
significant feature of an RRS process is the retardation effect between 
the two steps of scattering (see Fig. \ref{RRS}),
which is completely absent in the nonresonant theory \cite{wang02_nrs}. 
Such retardation effects can be studied
within the FL model in all dimensions \cite{wang02_rrs}
or within the LL model in the one-dimensional system \cite{Wang00}. 
The polarized RRS spectra calculated in both models (FL and LL) 
are very similar
far from resonance: the main contribution is from the collective CDE plasmon
mode (or charge boson) excitation at plasmon energy, $\omega=\omega_\rho(q)$,
but a relatively small (but finite) single particle excitation
(or the singlet spinon excitation in LL model) can also appear at energy,
$\omega=|q|v_F$. However, close to the resonance, the results calculated by 
these two models are quite different: The SPE weight in the FL model can 
become comparable to the collective CDE mode weight, whereas the weight of SSE
in the LL model is always smaller than the charge boson weight 
\cite{endnote0,future}.

Therefore, a crucial question is the extent to which Raman scattering 
experiments reveal LL features
characteristic of the one dimensional physics. The differences between
a Luttinger liquid and a Fermi liquid are most evident 
in the single electron problem,
which is measurable in principle by photoemission
\cite{photoemission} or tunneling \cite{tunneling} experiments but 
is unfortunately not directly measurable by {\it nonresonant} Raman
scattering, which involves the creation of particle-hole 
pairs in the conduction band. 
However, the situation can be different when considering the resonance 
feature explicitly, because only single electron (not charge or spin bosons) 
excitations between the conduction band and valence band are possible.
Therefore, from the perspective of single particle properties, 
we suggest that the RRS process near the resonance condition
can in principle be also a tool to experimentally distinguish the Luttinger 
liquid behavior from the Fermi liquid behavior.
We note that far away from resonance,
the photon frequency dependence of the spectral weight of the CDE mode 
is different in these two models: it scales as
$|\Omega|^{-2}$ in the FL theory, but decreases slower as
$|\Omega|^{2\alpha-2}$ in the LL theory \cite{endnote}.

The LL calculations predict a much smaller relative
spectral weight in the SPE mode compared with the CDE mode
than that observed in the existing
experiments. This disagrees with the conclusion in 
Ref. [\onlinecite{Sassetti98}],
where the resonance matrix elements are not self-consistently treated
in the Luttinger liquid theory \cite{Wang00}. 
It is possible that the experiments are not yet probing 
the low energy limit where
the Luttinger liquid model is fully applicable. This can be attributed to,
for example, the
finite size effects of the wire, finite band curvature for 
excitations about the Fermi surface, and/or finite temperature cut-off, etc.
All of which may suppress the LL features in the experiment making 
it indistinguishable from the FL theory results.

\section{Summary}
\label{summary}

In this paper, we review the various theories for the
resonant Raman scattering experiment, a powerful tool to
study the elementary electronic excitations in low-dimensional
semiconductor structures. In addition to the known 
collective plasmon
(charge boson) excitations, the Luttinger liquid can in principle
have an additional singlet spinon excitation which could mimic 
the single particle excitation behavior in the Fermi liquid model.
The polarized RRS spectra calculated in the FL model
and the LL model, however, are different in the relative weights of 
these excitations with the FL model in general showing much better 
agreement with experiment.
This may be because the strong interband scattering invariably present in the 
resonant process 
mixes the conduction band and the valence band states together, leading to 
an ``imperfect'' one-dimensional system for electron excitations
near the conduction band Fermi surface 
(i.e. electrons can be excited from below the conduction band Fermi surface 
to above the Fermi surface via the mediation of the valence band). 
However, in our present theory, we do not include excitonic
effects (interaction between conduction band electrons and valence band
holes) in calculation, which might be crucial during the Raman
scattering process near resonance conditions. Therefore,
%
%
Further theoretical and experimental studies are required for the unambiguous 
demonstration of the Luttinger liquid behavior in the RRS spectra 
of the semiconductor quantum wire structures.
This somewhat unclear RRS situation, where the Fermi liquid model
seems to produce apparent better quantitative agreement with the 
experimental observations in GaAs quantum wires, is in sharp
contrast with the tunneling spectroscopic transport studies of GaAs 
quantum wires \cite{tunneling} which are well-explained by the Luttinger
liquid theory \cite{bert}.

This review is a brief summary of our recent works in the Raman scattering 
theory of one-dimensional electronic systems 
(Refs. [\onlinecite{wang02_nrs,Dassarma99,Wang00,wang02_rrs,future}]).
Readers can find more details and references therein.



\thebibliography{}

\bibitem{technique}
See, for example,
Y.C. Chang, L.L. Chang, L. Esaki,
Appl. Phys. Lett. {\bf 47} 1324 (1985); 
A.R. Go$\tilde{\rm n}$i, K.W. West, A. Pinczuk, 
H.U. Baranger, H.L. Stormer, Appl. Phys. Lett. {\bf 61} 1956 (1992).

\bibitem{Lai}
W.Y. Lai and S. Das Sarma, Phys. Rev. B {\bf 33}, 8874 (1986).

\bibitem{effective_mass2} Q. P. Li and S. Das Sarma,
           Phys. Rev. B, \textbf{43}, 11768 (1991);
           Q. P. Li, S. Das Sarma, and R. Joynt, \textbf{45} 13713 (1992)

\bibitem{effective_mass3}
S. Das Sarma and D.-W. Wang
Phys. Rev. Lett. {\bf 84}, 2010 (2000);
D.-W. Wang and S. Das Sarma
Phys. Rev. B {\bf 64}, 195313 (2001).

\bibitem{dassarma_review}
S. Das Sarma, Elementary Excitations in Low-Dimensional 
Semiconductor Structures.
p. 499 in \textit{Light scattering in Semiconductor Structures
and Superlattices}, edited by D.J. Lockwood and J.F. Young
(Plenum, New York, 1991).

\bibitem{ll}
S. Tomonaga, Prog. Theor. Phys. {\bf 5}, 544 (1950);
J.M. Luttinger, J. Math, Phys. N.Y. {\bf 4}, 1154 (1963);
F. D. M. Haldane, J. Phys. C, \textbf{14}, 2585 (1981).

\bibitem{voit95} J. Voit,
                 Rep. Prog. Phys. \textbf{58}, 977 (1995).

\bibitem{schulzemery}
 H. J. Schulz, in \textit{Correlated Electron Systems}, edited by V. J. Emery
   (World Scientific, Singapore 1993).

\bibitem{manhan}
 G. D. Manhan, \textit{Many Particle Physics}, (Plenum,
            New York, 1990).

\bibitem{pinczuk} A. Pinczuk, B.S. Dennis, L.N. Pfeiffer, and K.W. West,
              Philosophical Magazine B {\bf 70}, 429 (1994) and 
              references therein.

\bibitem{pinczuk_1D} 
A. Schmeller, A.R. Goñi, A. Pinczuk, J.S. Weiner, 
J.M. Calleja, B.S. Dennis, L.N. Pfeiffer, and K.W. West,
Phys. Rev. B {\bf 49}, 14778 (1994).

\bibitem{pinczuk_2D}
J.E. Zucker, A. Pinczuk, D.S. Chemla, and A.C. Gossard,
Phys. Rev. B {\bf 35}, 2892 (1987);
G. Danan, A. Pinczuk, J.P. Valladares, L.N. Pfeiffer, K.W. West, and C.W. Tu,
Phys. Rev. B 39, 5512-5515 (1989).

\bibitem{QH_exp}
I. Dujovne, A. Pinczuk, M. Kang, B.S. Dennis, L.N. Pfeiffer, and K.W. West
Phys. Rev. Lett. {\bf 90}, 036803 (2003);
A. Pinczuk, J.P. Valladares, D. Heiman, A.C. Gossard, J.H. English, 
C.W. Tu, L. Pfeiffer, and K. West
Phys. Rev. Lett. {\bf 61}, 2701 (1988);
A. Pinczuk, B. S. Dennis, L. N. Pfeiffer, and K. West
Phys. Rev. Lett. {\bf 70}, 3983 (1993).

\bibitem{QWR_ref} A. R. Go$\tilde{\mathrm{n}}$i, A. Pinczuk, J. S. Weiner,
               J. M. Calleja, B. S. Dennis, L. N. Pfeiffer, and K. W. West,
                      Phys. Rev. Lett. \textbf{67}, 3298 (1991).

\bibitem{rrs_exp96} C. Sch$\ddot{\rm{u}}$ller, G. Biese, K. Keller,
                    C. Steinebach, and D. Heitmann,
                      Phys. Rev. B \textbf{54}, R17304 (1996).

\bibitem{pines} D. Pines, and P. Nozieres, 
              \textit{The Theory of Quantum Liquids}
              (Benjamin, New York, 1966).

\bibitem{3D} J. K. Jain, and P. B. Allen,
              Phys. Rev. Lett. \textbf{54}, 947 (1985);
             J. K. Jain, and P. B. Allen,
              Phys. Rev. Lett. \textbf{54}, 2437 (1985);
             S. Das Sarma and E. H. Hwang,
              Phys. Rev. Lett. \textbf{81}, 4216 (1998).

\bibitem{2D} J. K. Jain, and S. Das Sarma,
              Phys. Rev. B \textbf{36}, 5949 (1987);
             J. K. Jain, and S. Das Sarma,
              \textit{Surf. Sci.} \textbf{196}, 466 (1988);
             J. K. Jain and P. B. Allen, Phys. Rev. Lett.
              \textbf{54}, 947 (1985);
 S. Das Sarma and P. Tamborenia, Phys. Rev. Lett, \textbf{73}, 1971 (1994).

\bibitem{wang02_nrs} D.W. Wang and S. Das Sarma,
             Phys. Rev. B {\bf 65}, 035103 (2002).

\bibitem{Jusserand00} B. Jusserand, M. N. Vijayaraghavan, F. Laruelle,
             A. Cavanna, and B. Etienne,
             Phys. Rev. Lett. \textbf{85}, 5400 (2000).

\bibitem{new_data}
J. Rubio, J.M. Calleja, A. Pinczuk, B.S. Dennis, L.N. Pfeiffer, K.W. West,
Solid State Commun. {\bf 125} 149-153 (2003).

\bibitem{Sassetti98} M. Sassetti and B. Kramer, Phys. Rev. Lett. \textbf{80},
              1485 (1998).

\bibitem{Dassarma99} S. Das Sarma and D.-W. Wang,
             Phys. Rev. Lett. \textbf{83}, 816 (1999).

\bibitem{Walf} P. A. Wolff,
             Phys. Rev. Lett. \textbf{16}, 225 (1966);
              Phys. Rev. \textbf{171}, 436 (1968);
             F. A. Blum,
              Phys. Rev. B \textbf{1}, 1125 (1970).]

\bibitem{sakurai} J. J. Sakurai, \textit{Advanced Quantum Mechanics}
                 (Addison-Wesley, Redwood, 1984).

\bibitem{fetter}A. L. Fetter and J. D. Walecka,
         \textit{Quantum Theory of Many-particle
         Systems}, (McGraw-Hill, San Francisco, 1971).

\bibitem{larkin}
I.E. Dzyaloshinsky and A.I. Larkin, Zh. eksp. teor. Fiz. {\bf 65}, 411 (1973)
(English translation: Soviet Phys. JETP {\bf 38}, 202 (1974)).

\bibitem{hubbard}
 J. Hubbard, Proc. R. Soc. London Ser. A \textbf{243}, 336 (1957).

\bibitem{schulz93}
H.J. Schulz, Phys. Rev. Lett. {\bf 71}, 1864 (1993).

\bibitem{lieb}
  E. H. Lieb and F. Y. Wu, Phys. Rev. Lett. \textbf{20}, 1445 (1968).

\bibitem{coll}
 C. F. Coll, Phys. Rev. B \textbf{9}, 2150 (1974).

\bibitem{yang}
  C. N. Yang, Phys. Rev. Lett. \textbf{19}, 1312 (1967).

\bibitem{gagliono}
 E. R. Gagliono, E. Dagotto and A. Moreo, 
Phys. Rev. B \textbf{34}, 1677 (1986);
 E. R. Gagliono and C. A. Balseiro, Phys. Rev. Lett. \textbf{59}, 2999
     (1987);
 E. R. Gagliono and C. A. Balseiro, Phys. Rev. B \textbf{38}, 11766 (1988);
 E. R. Gagliono and C. A. Balseiro, Phys. Rev. Lett. \textbf{62}, 1154 (1989).

\bibitem{num_sc_separation}
 R. Preuss, A. Muramatsu, W. von der Linden, F. F. Assaad, and W. Hanke,
    Phys. Rev. Lett. \textbf{73}, 732 (1994);
 M. G. Zacher, E. Arrigoni, W. Hanke and J. R. Schrieffer,
    Phys. Rev. B \textbf{57}, 6370 (1998).

\bibitem{Wang00} D.-W. Wang, A. J. Millis, and S. Das Sarma,
             Phys. Rev. Lett. \textbf{85}, 4570 (2000).

\bibitem{wang02_rrs} D.-W. Wang and S. Das Sarma,
            Phys. Rev. B {\bf 65}, 125322 (2002).

\bibitem{endnote0} However, we note that the SPE spectral weight near resonance
depend on the actual value of $\lambda$ in the present FL theory.
It actually diverges when $\lambda$ is taken to be zero. Details is
discussed in Ref. \cite{future}.

\bibitem{future} D.-W. Wang, A. J. Millis, and S. Das Sarma, cond-mat/0405452.

\bibitem{endnote}
We note that the Luttinger liquid theory is an effective low energy
theory about the Fermi surface. Therefore in principle it 
is not appropriate for off resonance region, $\Omega>E_F$. But
the resonance energy dependence may be still observable within the
existing considtions of RRS experiments, say $0.1 E_F<\Omega<E_F$.

\bibitem{photoemission}
H. W. Yeom, K. Horikoshi, H. M. Zhang, K. Ono, and R. I. G. Uhrberg,
Phys. Rev. B {\bf 65}, 241307 (2002);
T. Mizokawa, K. Nakada, C. Kim, Z.-X. Shen, T. Yoshida, 
A. Fujimori, S. Horii, Y. Yamada, H. Ikuta, and U. Mizutani,
Phys. Rev. B {\bf 65}, 193101 (2002).

\bibitem{tunneling}
O.M. Auslaender, A.Yacoby, R. De Picciotto, 
K.W. Baldwin, L.N. Pfeiffer, K.W. West, Science {\bf 295}, 825 (2002).

\bibitem{bert}
Y. Tserkovnyak, B.I. Halperin, O.M. Auslaender, and A.Yacoby,
Phys. Rev. Lett. \textbf{89}, 136805 (2002);
Phys. Rev. B \textbf{68}, 125312 (2003).
\end{document}